\documentclass[prb,aps,twocolumn]{revtex4}

\usepackage{graphicx}
\usepackage{amsmath}
\usepackage{amssymb}
\usepackage{ifthen}
\usepackage{booktabs}

\usepackage{graphicx}
\usepackage{dcolumn}
\usepackage{bm}
\usepackage{longtable}
\usepackage{gensymb}

\newcommand{\e}{\varepsilon}

\renewcommand{\(}{\left(}
\renewcommand{\)}{\right)}
\renewcommand{\vec}[1]{\mathbf{#1}}

\newcommand{\dif}[2]{\frac{{\rm d}#1}{{{\rm d}#2}}}
\newcommand{\pdif}[2]{\frac{{\rm \partial}#1}{{{\rm \partial}#2}}}

\begin{document}

\title{Two-dimensional colloidal fluids exhibiting pattern formation}
\author{Blesson Chacko} 
\author{Christopher Chalmers}
\author{Andrew J. Archer}
\affiliation{Department of Mathematical Sciences, Loughborough University, Loughborough, Leicestershire, LE11 3TU, UK}

\begin{abstract}
Fluids with competing short range attraction and long range repulsive interactions between the particles can exhibit a variety of microphase separated structures. We develop a lattice-gas (generalised Ising) model and analyse the phase diagram using Monte Carlo computer simulations and also with density functional theory (DFT). The DFT predictions for the structures formed are in good agreement with the results from the simulations, which occur in the portion of the phase diagram where the theory predicts the uniform fluid to be linearly unstable. However, the mean-field DFT does not correctly describe the transitions between the different morphologies, which the simulations show to be analogous to micelle formation. We determine how the heat capacity varies as the model parameters are changed. There are peaks in the heat capacity at state points where the morphology changes occur. We also map the lattice model onto a continuum DFT that facilitates a simplification of the stability analysis of the uniform fluid.
\end{abstract}

\maketitle

\section{Introduction}
\label{sec:intro}

When the forces between colloidal particles suspended in a liquid are sufficiently strongly attractive, they can exhibit phase separation into a high density colloidal fluid, referred as a colloidal ``liquid" and low density suspension, a colloidal ``gas".\cite{Hansen:2013aa} However, in some circumstances, the interactions can be attractive at short ranges when the particle cores are close to one another, but at longer ranges be repulsive. These short-range attractive, long-range repulsive (SALR) potentials can arise in certain suspensions of charged colloids and polymers \cite{CADB05} and also in protein solutions.\cite{Stradner:2004aa} Self-consistent Ornstein-Zernike approximation (SCOZA) integral equation theory for a model of such systems,\cite{Pini:2000aa, Pini:2006aa} showed that when the long range repulsion is not too strong there is a large region of the phase diagram where the correlations in the fluid show significant fluctuation effects and where the compressibility increases significantly. The SCOZA theory (which is sophisticated and rather accurate) was also compared with results from DFT,\cite{Archer:2007ab} which showed good agreement between the theories for the liquid structure. When the long range repulsion is further increased, the SALR interaction between the particles gives rise to pattern formation in the fluid state, such as gathering to form clusters, stripes (lamellas) and holes (bubbles), referred to as microphase separation. In Ref.~\onlinecite{Archer:2007aa} Monte Carlo (MC) computer simulations and integral equation theory was used to understand the details of the relation between the liquid-vapour transition line and the occurrence of any microphase separated phases. As the repulsion strength is increased, starting from the critical point, the gas-liquid phase separation is replaced by microphase separation. In Ref.~\onlinecite{SFL14}, a study of the cluster formation showed that it is very similar to micelle formation in aqueous surfactant solutions. {However, for the system considered in Ref.~\onlinecite{bomont2012communication}, discontinuities in thermodynamic quantities were observed at the onset of cluster formation, suggesting it is indeed a phase transition.}

Further understanding of the phase ordering in SALR systems was {recently} gained by Pekalski and co-workers\cite{Pekalski:2013aa} by studying a simple one-dimensional lattice model, in which the SALR interaction was modelled using an attractive interaction between neighbouring particles, repulsion between the third neighbours and no interaction between second neighbours or any other neighbours. An exact solution was presented using the transfer matrix method. The same SALR system was then extended to two-dimensions (2D) on a triangular lattice,\cite{Pekalski:2014ab, Almarza:2014aa} where microphase separated phases and also a reentrant uniform liquid is observed in the phase diagram. {This approach, based on using lattice models to elucidate the nature of the structure formation in systems with competing interactions, has a long track record, going back to seminal works, such as Refs.\ \onlinecite{selke1980two, landau1985phase}. There are several advantages of using lattice models stemming from the fact that they are much more straightforward to analyse than the equivalent continuum models and also the computations are much simpler, allowing larger systems to be simulated over longer times. Due to the fact that the clusters and other structures formed can be more than an order of magnitude larger than the size of the individual particles, to properly observe the microphase formation, the system size generally needs to be much larger than that one would use for studying simple gas-liquid systems.} There have also been other (field) theoretical and simulation studies considering aspects of the phase behaviour of a variety of fluids interacting via SALR potentials.\cite{Andelman:1986aa,Nussinov:1999aa,Muratov:2002aa,Tarzia:2007aa}

The {more recent} interest in SALR systems in 2D stems from the experimental observation of microphase-ordering of nanoparticles at a water-air interface,\cite{Sear:1999aa,Sear:1999ab} which led to theoretical and simulation work to understand the nature of the structures that are formed. Imperio and Reatto\cite{Imperio:2004aa,Imperio:2006aa,Imperio:2007aa} made a detailed study of the phase diagram using parallel-tempering MC simulations to determine the location in the phase diagram of the microphase separated states for a 2D fluid of particles interacting via the double-exponential pair potential
\begin{equation}
	u(r) = \left\{
		\begin{array}{l l}
			\infty ,
			& \mathrm{if}  \quad {r < \sigma}
		\\[0.5em]
			-\cfrac{\e_a\sigma^2}{R_a^2} e^{-r/R_a}
			+ \cfrac{\e_r\sigma^2}{R_r^2} e^{-r/R_r},
			& \mathrm{otherwise}
		\end{array}
	\right.
\end{equation}
where $r$ is the distance between the centres of the particles, which have a hard-core of diameter $\sigma$. The short range attraction has strength determined by $\e_a$ and range $R_a$. Similarly, the repulsion strength is determined by $\e_r$ and has range $R_r$. When $R_a = \sigma$, $R_r = 2\sigma$ and $\e_a = \e_r = \e$ , microphase ordering is observed for temperatures $k_BT/\e \lesssim 0.6$, where $k_B$ is Boltzmann's constant. At lower densities this takes the form of clusters or ``droplets", whilst at higher densities striped structures were observed. At even higher densities a hole phase is observed, although here the simulations can be difficult to perform. Imperio and Reatto \cite{Imperio:2004aa,Imperio:2006aa,Imperio:2007aa} showed that at the onset of microphase ordering, one observes a peak in the heat capacity and this was used to identify the location in the phase diagram of the microphase ordered states. Following this, a DFT model for this system was developed,\cite{Archer:2008aa} which is in good qualitative agreement with simulation results with regard to the topology of the phase diagram and the structure of the fluid and inhomogeneous phases. The DFT also predicts that the transitions from the uniform to the modulated fluid phases are all either first or second order phase transitions,\cite{Archer:2008aa} However, the DFT is a mean-field theory and so one should be cautious about accepting this prediction of the theory.

The aim of the work described here is to study the formation of patterns using both MC computer simulations and also DFT for a 2D lattice model in order to determine the nature of the transitions to and between the different microphase ordered structures and also to compare between the methods in order to elucidate what aspects of the microphase ordering the mean-field DFT is able to describe. We fix the strength of the repulsion between the particles to a particular value and we also fix the temperature and then calculate the properties of the fluid as the density and the strength of the attractive interactions between the particles are varied. In particular, we calculate the heat capacity and determine the phase diagram. We also map the lattice model onto a continuum DFT that allows a simple calculation of roughly where in the phase diagram one can expect to find the microphase ordering. This takes the form of a linear stability analysis.

This paper is laid out as follows: In Sec.~\ref{sec:model} we define the model fluid and in Sec.\ \ref{sec:MC} we present MC computer simulation results, including for the heat capacity, {for the ratio of particles in the system within the clusters as the total density in the system is increased and for the static structure factor}. In Sec.\ \ref{sec:Lattice_DFT} we present the lattice DFT results, comparing with the MC results and calculating the fluid phase diagram. In Sec.\ \ref{sec:LS} we map onto a continuum DFT and discuss the linear stability of the fluid. Finally, in Sec.\ \ref{sec:conc} we draw our conclusions.

\section{The model fluid}
\label{sec:model}

\begin{figure}[b]
  \centering
  \vspace*{1em}
  \includegraphics[width=0.45\textwidth]{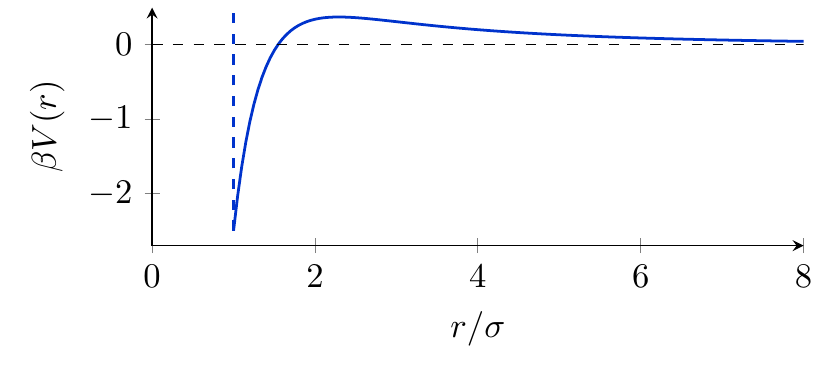} 
  \caption[Double-Yukawa pair potential]{The double-Yukawa pair interaction potential between the particles, in the case when the parameters are $\beta A=1.5$, $z_1=2$, $z_2=0.2$ and $\beta\e=4$.}
  \label{fig:potential}
\end{figure}

We assume that the colloids interact via the pair potential
\begin{equation}
    u(r) = \left\{
           \begin{array}{l l}
             V(r) 
               & \quad  {r \ge \sigma}
             \\
             \infty
               & \quad {r < \sigma},
           \end{array}
         \right.
       \label{eq:DY}
\end{equation}
where $r$ is the distance between the centres of the two particles and the tail of the potential is given by the double-Yukawa potential\cite{Archer:2007aa}
\begin{equation}
    V(r) = \left\{
           \begin{array}{l l}
             -\cfrac{ \e e^{-z_1(r-\sigma)/\sigma}}{r/\sigma} + \cfrac{A e^{-z_2(r-\sigma)/\sigma} }{r/\sigma} 
               & \quad  {r \ge \sigma}
             \\
             0
               & \quad {r < \sigma}
           \end{array}
         \right.
       \label{eq:tail}
\end{equation}
where $\e$ is the attraction strength coefficient and $A$ is the repulsion strength coefficient. The parameters $z_1$ and $z_2$ determine the range of the attraction and repulsion, respectively. $\sigma$ is the diameter of the particles, which we set to be our unit of length. We fix the coefficients $z_1=2$ and $z_2=0.2$ so that the potential is of the form illustrated in Fig.~\ref{fig:potential}.

In order to simplify the analysis {and to reduce the computational costs}, we assume that the positions of the particles are discrete variables, and represent the fluid via a 2D lattice model, containing $M$ lattice sites and with periodic boundary conditions. We use a square lattice {of size $L\times L$}, with lattice spacing equal to the diameter of the particles $\sigma$ and we assume that each lattice site can be occupied by at most one colloid. We denote a particular configuration of particles by a set of occupation numbers $\{n_{i}\}$, such that, if the site $i$ is empty, then $n_{i}=0$ and  $n_{i}=1$, if it is occupied. Note that $i$ here is used as a short form for the position on the 2D lattice, at point $(j,k)$. We treat the system in the grand canonical ensemble and so the Hamiltonian of our lattice model can be written as \cite{Hughes:2013aa}
\begin{equation}
  E\(\{n_{i}\}\) = \sum_{i=1}^{M}n_{i}(\Phi_{i}-\mu)
                    +\frac{1}{2}\sum_{i,j}V_{i,j}n_{i}n_{j}\, ,
  \label{eq:Energy_config}
\end{equation}
where $\Phi_{i}$ is the external potential at the lattice site $i$ and $\mu$ is the chemical potential which determines the number of particles in the system $N$. The final term is the energy contribution due to the interactions between particles, where $V_{i,j}$ is the pair interaction potential between two particles at sites $i$ and $j$, which is the discrete lattice version of the potential in Eq.~\eqref{eq:tail}, i.e.~evaluated by taking $r$ in Eq.~\eqref{eq:tail} to be the distance between sites $i$ and $j$. We also assume that there are no three-body or higher-body interactions between the particles. Since here we only consider the ordering in the bulk fluid, we henceforth assume that $\Phi_i = 0$, $\forall i$. Also, in all our MC and DFT results below, we truncate the tail of the pair potential beyond $r=r_c=16\sigma$. {It is also worth noting that the lattice model Hamiltonian \eqref{eq:Energy_config} has a symmetry between particles and holes (i.e.\ replacing $n_i \to 1-n_i$) that, as we show below, results in the phase diagram of the system being symmetric around the density $\rho=\langle n_i \rangle = 1/2$.}

\section{Monte Carlo}
\label{sec:MC}

\begin{figure*}[tb]
  \centering
  \includegraphics[width=0.98\textwidth]{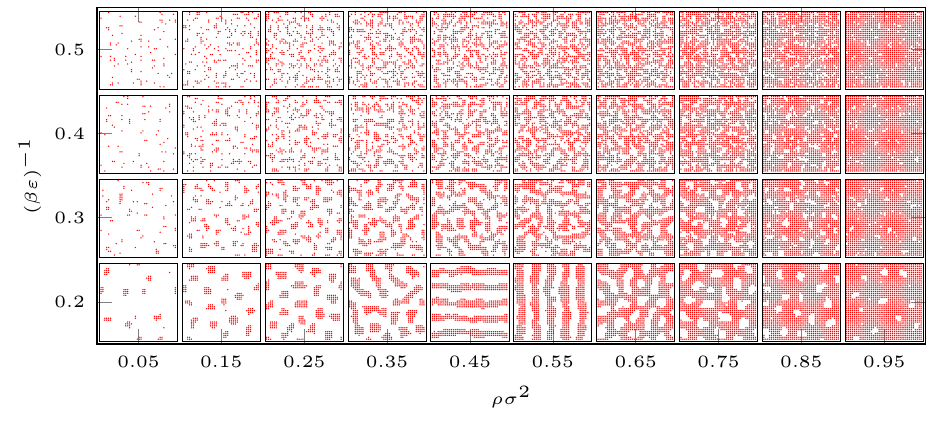} 
  \caption[Monte Carlo simulation snapshots for different parameters]
  {Snapshots of typical configurations for a $40\sigma\times40\sigma$ size system with $\beta A = 1.5$, $z_1=2$ and $z_2=0.2$, obtained from grand-canonical MC simulations for various values of the average density and varying values of $(\beta\e)^{-1}$.}
  \label{fig:mc_snapshots}
\end{figure*}

We study the system using standard Metropolis MC simulations.\cite{Landau:Binder} The lattice is {initiated} in a state where all the sites are randomly occupied by a particle with probability 0.5. At each step during the simulation, a random lattice site $i$ is selected and we then calculate the change in energy $\Delta E$ using Eq.~\eqref{eq:Energy_config} when the occupation number for that lattice site is replaced $n_i\to(1-n_i)$. Thus, if the site is already occupied, the trial change is to remove the particle and if the site is unoccupied, the trial move it to insert a particle at that site. If $\Delta E$ is negative, then we keep the change. Otherwise, we only keep the change with probability, ${e^{-\beta \Delta E}}$.

In Fig.~\ref{fig:mc_snapshots}, we display typical snapshots from our MC simulations for a range of state points, for various average densities $\rho=\langle N \rangle/M$ (determined by the value of the chemical potential $\mu$) and several values of the inverse attraction strength parameter, $(\beta\e)^{-1}$. At low values of $(\beta\e)^{-1}$, as the average density is increased, the system exhibits a sequence of microphase separated structures. At very low densities, the system forms a gas phase. Increasing $\rho$, when the value of $(\beta\e)^{-1}$ is low enough, we see the particles are arranged into clusters of a characteristic size. Further increasing $\rho$, we observe stripe like patterns for $\rho\sigma^2 \sim 0.5$. At even higher densities, we observe a fluid containing `bubbles', again with a characteristic size. Finally, for large $\rho$, the system is almost entirely full of particles, forming a dense liquid.  Increasing $(\beta\e)^{-1}$ leads to the particles becoming less correlated, making it difficult to identify what microphase separation occurs, if any. 

\subsection{Heat Capacity}
\label{sec:heat_capacity}

We calculate the heat capacity as the chemical potential $\mu$ is varied, in order to identity the regions of the phase diagram where the microphase separation occurs. At a phase transition, in the thermodynamic limit, there is normally either a discontinuity or a divergence in the heat capacity. For finite size systems, these show up as peaks in the heat capacity. Recall also that a ``bump" in the heat capacity was observed at the onset of microphase ordering in the simulations of Imperio and Reatto.\cite{Imperio:2006aa} The heat capacity at constant volume can be obtained from the following derivative with respect to temperature,\cite{Bowley:1999aa}
\begin{equation}
  C_V = \(\pdif{U}{T}\)_V,
\end{equation}
where the internal energy $U=\langle E \rangle$. Alternatively, it can be calculated by measuring the energy fluctuations within the system,\cite{Chandler:1987aa}
\begin{equation}
   C_V = \frac{ \langle E^2 \rangle - \langle E \rangle^2 }{k_B T^2}.
   \label{eq:C_v}
\end{equation}

\begin{figure}[!htbp]
  \centering
  \includegraphics[width=0.45\textwidth]{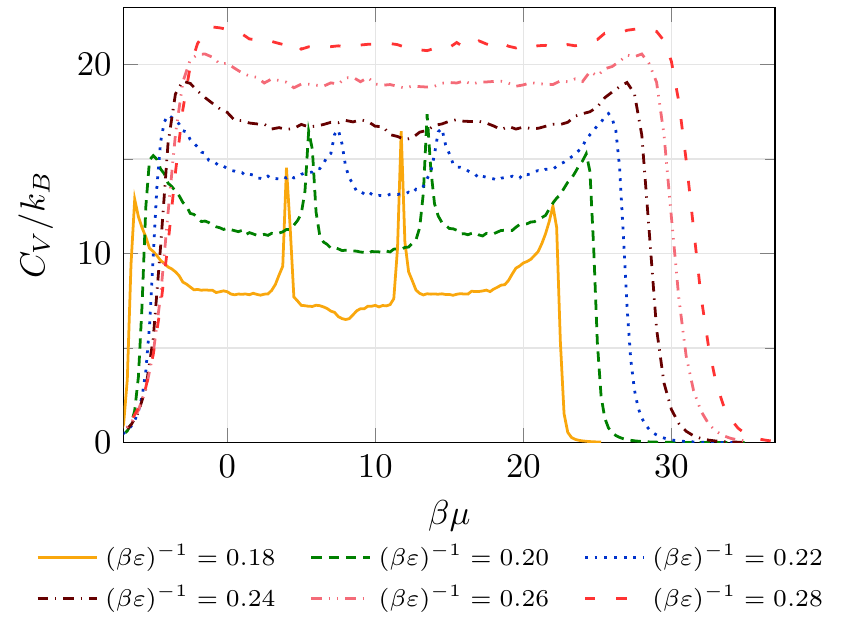} 
  \caption[Heat capacity against chemical potential from Monte Carlo simulations]
  	{Heat capacity verses chemical potential, $\mu$, for different values of $(\beta\e)^{-1}$, obtained from Monte Carlo simulations for a $40\sigma\times40\sigma$ size system with $\beta A = 1.5$, $z_1=2$ and $z_2=0.2$.}
  \label{fig:heat_capacity}
\end{figure}

A plot of the heat capacity of a system of size $40\sigma\times40\sigma$ as a function of $\mu$ and for various values of $(\beta\e)^{-1}$ calculated via Eq.~\eqref{eq:C_v} is shown in Fig.~\ref{fig:heat_capacity}. The heat capacity tends to zero when the system is completely empty or fully filled. This is as expected, since the system contains hardly any particles to give rise to energy fluctuations at lower values of the chemical potential, $\mu \to -\infty$, and in the opposite limit $\mu \to \infty$, the system is almost completely full of particles, so that the energy of the system, $E$, also does not fluctuate much in value.

For higher values of $(\beta\e)^{-1}$, we see in Fig.~\ref{fig:heat_capacity} that the heat capacity varies smoothly as $\mu$ is increased. However, for lower values of $(\beta\e)^{-1}$, we see four clear peaks in the heat capacity. These peaks correspond to changes in the structure of the fluid (see Fig.~\ref{fig:mc_snapshots}). Increasing $\mu$, the first peak corresponds to a change from a low density gas to a clustered structure. The second peak corresponds to the change from the cluster to the stripe morphology. The third peak to the change from stripe to bubble and then the final fourth peak to the change from a liquid containing bubbles to a dense liquid without bubbles. As $(\beta\e)^{-1}$ is increased, these peaks become smaller in height, eventually being so small that they cannot be identified.

The overall energy fluctuations in the system also get larger as one increases $(\beta\e)^{-1}$. The large (peak) values of the heat capacity $C_V$ corresponds to state points where there are large fluctuations in the energy of the system. Hence, the peak in $C_V$ identifies state points where there are multiple types of typical configurations, each with different energy $E$.

The presence of these peaks in the heat capacity at state points where the fluid changes morphology naturally leads to the question: are these phase transitions, or just changes in the nature of the fluid correlations? For the low density and high density peak, this question is addressed in the following section.

\subsection{Cluster Formation}
\label{sec:micellisation}

To answer the question just posed above: no, the cluster formation is not a phase transition, it is a continuous change analogous to micellisation in surfactants.

Recall that $N$ is the total number of particles in the system, which changes over time in a grand canonical system. We denote the average total number of particles to be $\langle{N}\rangle$, and $\langle{N_1}\rangle$ be the average number of particles that have no nearest or next nearest neighbours, which we refer to as ``lone particles". We also calculate the ratio of lone particles to the total number of particles, $R=\langle{N_1}\rangle/\langle{N}\rangle$, and how this quantity depends on the average density and chemical potential of the system.

\begin{figure*}[!htb]
  \centering
  \includegraphics[width=1\textwidth]{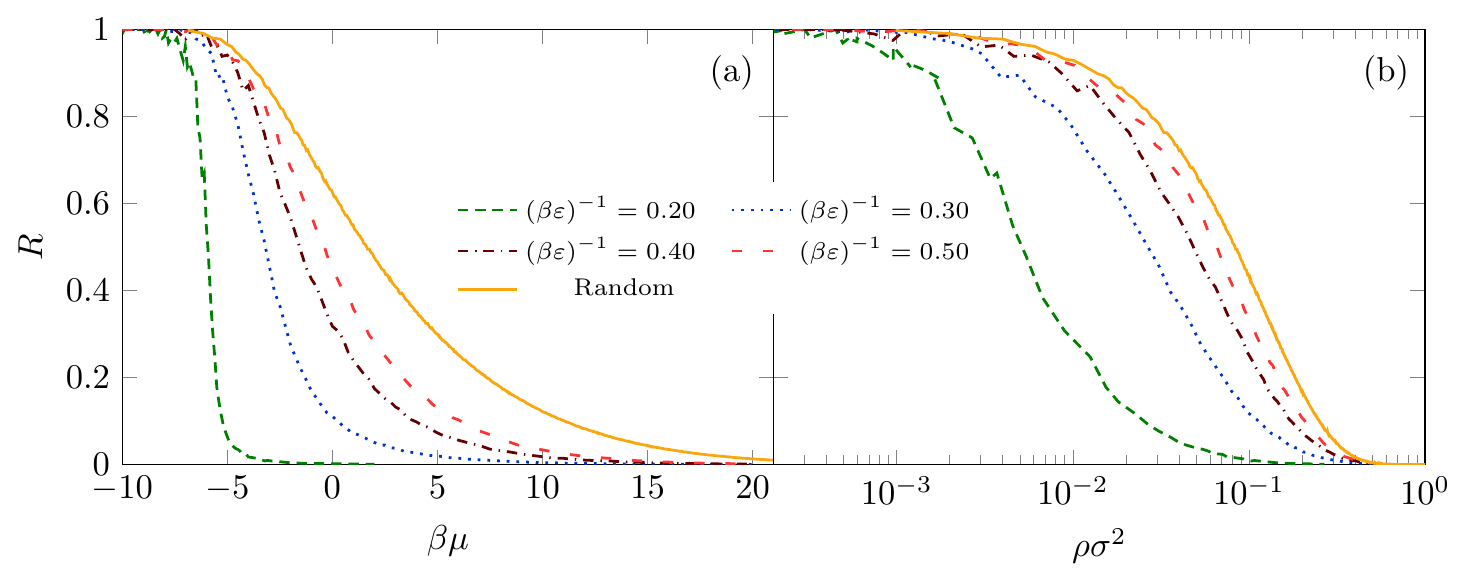} 
  \caption[Ratio of lone particles in the system against chemical potential and density]
  	{Ratio of lone particles in the system, $R$, for different values of $(\beta\e)^{-1}$, as a function of: (a) the chemical potential and (b) the average density. The solid line labelled ``Random" corresponds to the value of $R$ for the entirely random uncorrelated configurations that the system with $\e=0$ and $A=0$ exhibits. All other results are for the system with $\beta A = 1.5$, $z_1=2$ and $z_2=0.2$.}
  \label{fig:micelles}
\end{figure*}

In Fig.~\ref{fig:micelles}, we see that at lower values of chemical potential (i.e. low density), almost all the particles are lone particles and so $R \approx 1$. This is because when we have a small overall number of particles in the system, we are likely to find them all to be alone. As the attraction strength is increased (i.e. as $(\beta\e)^{-1}$ is decreased), we see that the drop in value from $R \approx 1$ for low $\mu$, to a value $R \ll 1$, becomes much steeper. For example, we see in Fig.~\ref{fig:micelles}(a) that when $(\beta\e)^{-1} = 0.2$, there is a very sudden drop in the value of $R$ at $\beta\mu \approx -5$. This corresponds to the change in morphology of the system from being mostly full of lone particles to the cluster phase. However, as can be seen in Fig.~\ref{fig:micelles}(b), where we display the variation of $R$ with the average density $\rho$ on a logarithmic scale, we see that actually the change in $R$ is continuous. The results in Fig.~\ref{fig:micelles} were calculated for a $40\sigma\times40\sigma$ size system, but these results do not change as the system size is increased (see also section~\ref{sec:MC:chang_box_size} below).

As we increase $(\beta\e)^{-1}$, we see the ratio of lone particles, $R$ tends towards the value that one would obtain for a system with $\e =0$ and $A=0$, i.e.\ where the particles are randomly distributed in the system. This is due to the decrease in particle correlations at higher values of $(\beta\e)^{-1}$, where the structure is essentially that of a highly supercritical fluid. Since the change in the ratio of lone particles is smooth and continuous as we increase the chemical potential (density) of the system, it is clear that the transition that we observe is not a phase transition, instead it is a structural change in the fluid much like micellisation at the critical micelle concentration (CMC).\cite{Israelachvili:2011aa}

Micellisation is the spontaneous self assembly of amphiphilic molecules in fluids. The forces that {hold} the amphiphiles together are generally weak, so that the structure within the micelles is fluid-like. Varying the solvent in which the micelles are suspended changes the interactions and so determines the structure and size of the micelles.\cite{Israelachvili:2011aa} The clusters we see are equivalent to spherical micelles, the bubbles are analogous to inverted micelles and the stripes to lamellar bilayer micelles. The similarities between the self-assembly of colloids and  amphiphilic molecules have been observed in many experimental, simulation and theoretical studies.\cite{ Archer:2007aa, Tarzia:2007aa, Imperio:2006aa,  Campbell:2005aa, Masri:2012aa} Indeed, Ciach and co-workers were able to describe both the SALR colloidal system and amphiphilic systems using {the} same functional,\cite{Ciach:2013aa} highlighting the many parallels between these systems.

{
Further support for the above conclusion about the nature of the structural changes in the system can be garnered from noting that the static structure factor $S(\vec{k})$ varies smoothly as $\mu$ is changed, taking the system from the low density gas state to the cluster morphology. $S(\vec{k})$ is a non-local quantity and so is sensitive to any onset of long range order, in contrast to $R$, which characterises only local (nearest neighbour) ordering. The static structure factor {we compute is}\cite{Hansen:2013aa,Imperio:2004aa}
\begin{align}
	S(\vec{k}) &= N^{-1}\left\langle \rho_\vec{k}\rho_\vec{-k} \right\rangle \nonumber \\
			     &= N^{-1} \Big\langle \big( \sum_{j=1}^{N} \cos( \vec{k}\cdot\vec{r}_j) \big)^2 
						 +\big( \sum_{j=1}^{N} \sin( \vec{k}\cdot\vec{r}_j) \big)^2 \Big\rangle ,
\end{align}
where $\rho_\vec{k} = \sum_{j=1}^{N} \exp( i\vec{k} \cdot \vec{r}_j)$, $N$ is the number of particles in the system, and $\vec{r}_j$ is the position on the lattice of each of the particle. In our calculations presented here, we fix the wavevector $\vec{k} = (k,0)$.

\begin{figure}[htbp]
  \centering
  \includegraphics[width=.49\textwidth]{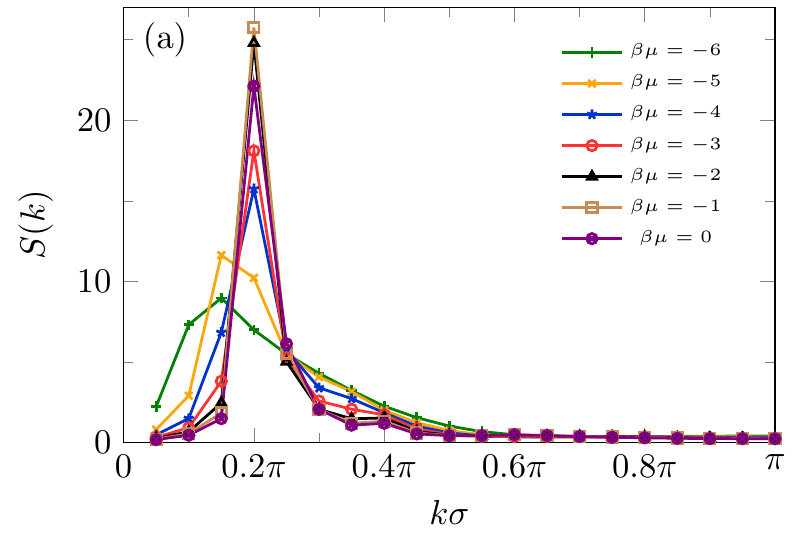}
  \includegraphics[width=.49\textwidth]{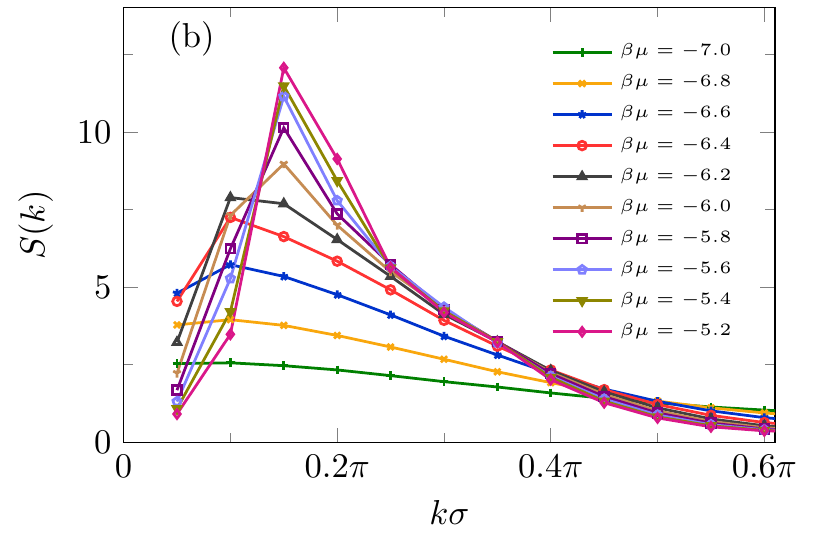}
  \caption{ {In (a) we display the static structure factor $S(k)$ for fixed $(\beta \e)^{-1} = 0.18$ and for a range of different values of the chemical potential $\mu$ where the cluster morphology is observed. The gas to cluster morphology change occurs at $\beta\mu\approx -6$, where there is a peak in the heat capacity (c.f.\ Fig.\ \ref{fig:heat_capacity}). In (b) we display $S(k)$ over a smaller range of values of $\mu$, going from the gas to the cluster morphology. We see that $S(k)$ varies smoothly as $\mu$ is varied -- see also Fig.\ \ref{fig:Sk_mu}.} }
  \label{fig:Sk_k}
\end{figure}

\begin{figure}[htbp]
  \centering
  \includegraphics[width=.49\textwidth]{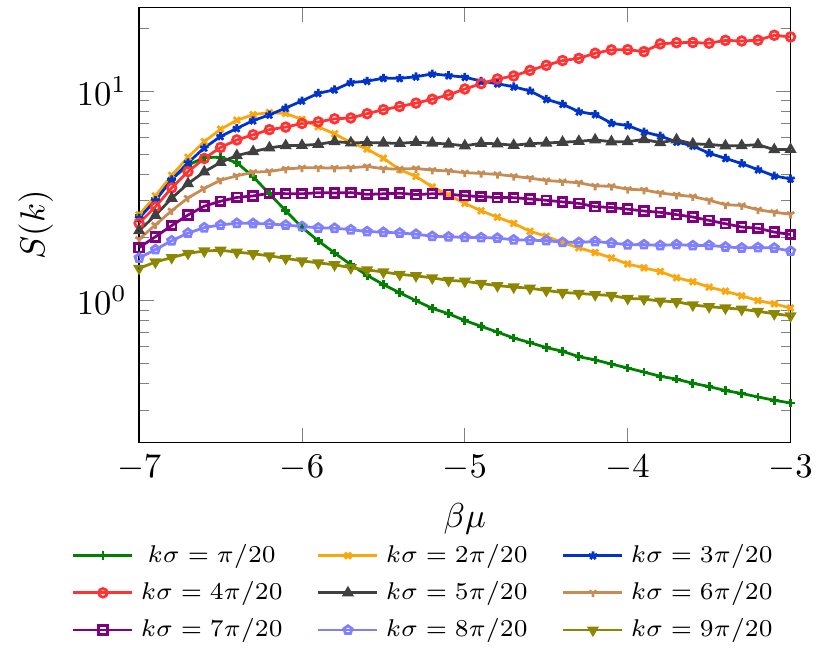}
  \caption{ {The static structure factor $S(k)$ for a range of different wavevectors $k$, as the chemical potential $\mu$ is varied, for fixed attraction strength $(\beta \e)^{-1} = 0.18$.} }
  \label{fig:Sk_mu}
\end{figure}

In Fig.~\ref{fig:Sk_k}(a) we display results for $S(k)$ for a range of state points where the cluster phase is observed, for fixed $(\beta \e)^{-1} = 0.18$. At lower densities (i.e.\ lower values of the chemical potential $\mu$), the peak in $S(k)$ is fairly broad with a maximum at $k\sigma=0.15\pi\approx0.47$, but for higher densities, the peak is sharper, with a maximum at $k\sigma=0.2\pi\approx0.63$. This is because at the higher densities the clusters interact more strongly with one another and the cluster-cluster correlations become significant. When $(\beta \e)^{-1} = 0.18$, the peak in the heat capacity for the gas to cluster transition occurs at $\beta\mu\approx -6$ [see Fig.~\ref{fig:heat_capacity}]. Fig.~\ref{fig:Sk_k}(b) shows that as $\mu$ is varied around this value, $S(k)$ varies smoothly, indicating there is no phase transition. This can also be seen from the plot in Fig.~\ref{fig:Sk_mu}, where we plot $S(k)$ for fixed values of $k$ as the chemical potential $\mu$ is varied, going from the low density gas state to deep in the region of the phase diagram where the cluster morphology occurs. One further interesting feature of the results in Fig.~\ref{fig:Sk_mu} is that in the cluster phase, the value of $S(k\sigma = \pi/4)$ is almost constant.}

{We also calculate the histogram of the probability of finding a given instantaneous density $\rho=N/M$ (not displayed). This has a single peak for all values of the chemical potential $\beta\mu\approx -6$, where the heat capacity peak occurs. This is in contrast to the three dimensional system considered in Ref.~\onlinecite{Archer:2007aa}, where a double peaked histogram is observed at the onset of cluster formation.}

\subsection{Changing Box Size}
\label{sec:MC:chang_box_size}

\begin{figure}[t]
	\centering
	\includegraphics[width=0.45\textwidth]{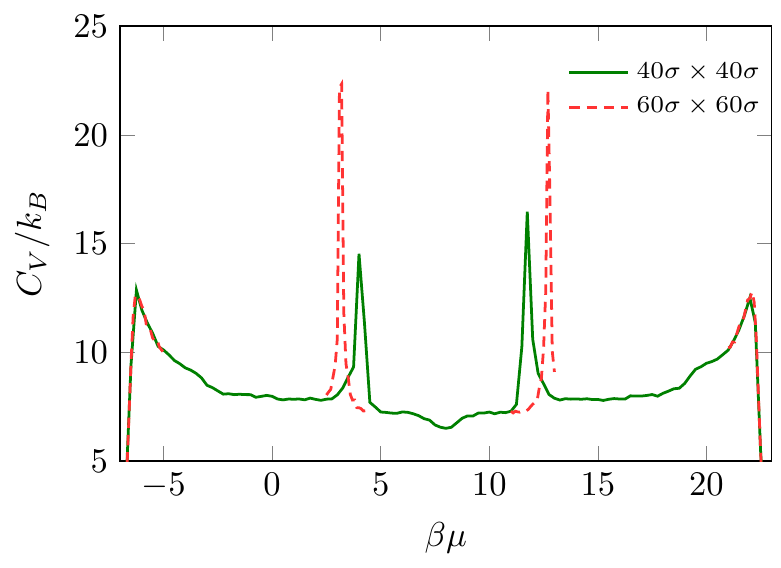} 
	\caption[The heat capacity versus chemical potential for two different box sizes]
  	{The heat capacity versus chemical potential $\mu$, for two different box sizes for $(\beta \e)^{-1} = 0.18$ [c.f.\ Fig.~\ref{fig:heat_capacity}].}
	\label{fig:heat_capacity_diff_box}
\end{figure}

Our MC simulations are performed in a finite size box with periodic boundary conditions to approximate an infinite system. However, for some of the transitions, it turns out that the box size is significant in determining the properties of the system. In Fig.~\ref{fig:heat_capacity_diff_box} we plot the heat capacity for $(\beta\e)^{-1} = 0.18$, calculated for simulations in a box of size $40\sigma\times40\sigma$ and compare with results for a larger box of size $60\sigma\times60\sigma$.

\begin{figure}[t]
	\centering
	\includegraphics[width=0.45\textwidth]{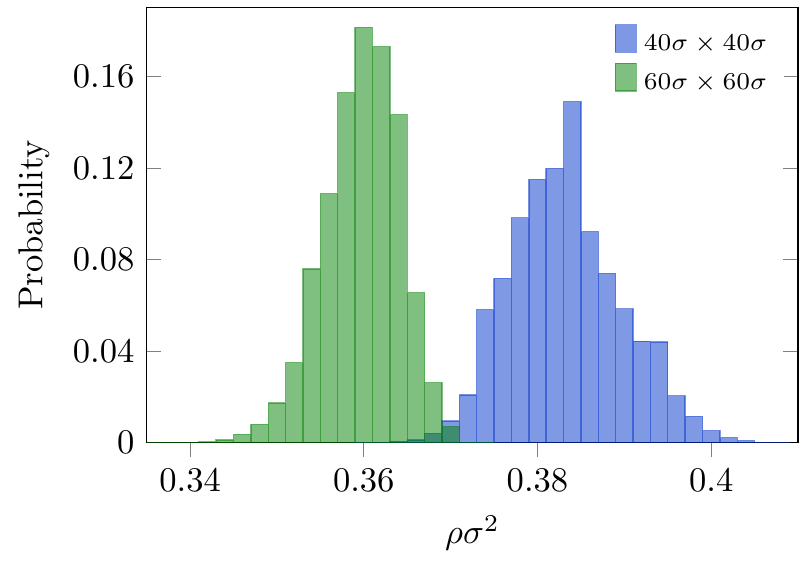} 
	\caption[Probability of finding a certain instantaneous density within a simulation]
  	{Probability of finding a certain instantaneous density $\rho=N/M$ calculated at the cluster to stripe transition (i.e.\ at the second peak in the heat capacity) for two different box size for $(\beta \e)^{-1} = 0.18$.}
	\label{fig:MC_prob_rho}
\end{figure}

\begin{figure}
	\centering
	\includegraphics[width=0.45\textwidth]{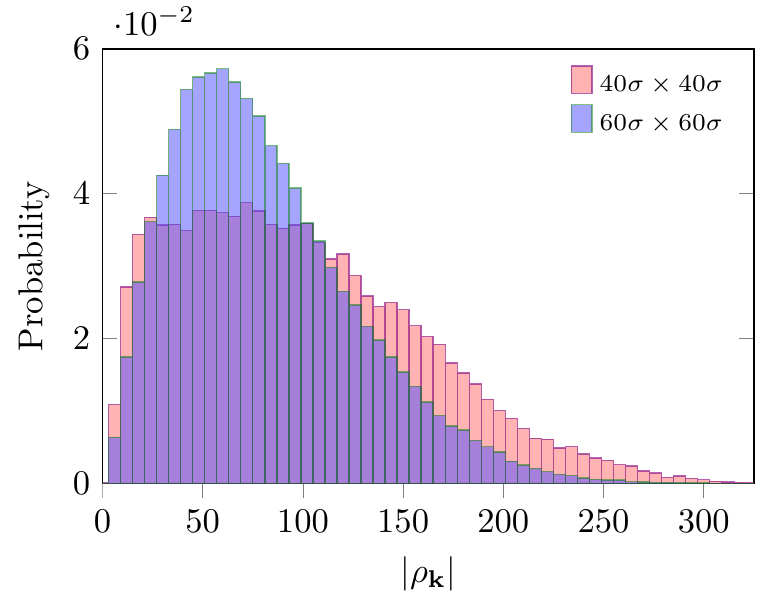} 
	\caption
	{{ Probability distribution for the density Fourier mode amplitude $|\rho_{\vec{k}}|$, with $k\sigma=0.2\pi$, calculated at the cluster to stripe transition (i.e.\ at the second peak in the heat capacity) for two different box sizes $L$ and for $(\beta \e)^{-1} = 0.18$.}}
	\label{fig:MC_Prob_Rho_k}
\end{figure}

In Fig.~\ref{fig:heat_capacity_diff_box}, we do not observe any effect of the finite box size on the value of the heat capacity at the peaks corresponding to the gas to cluster transition and also the bubbles to liquid transition. This confirms the conclusion in the previous section that this transition is akin to micellisation, and that there are no discernable  effects in the above results due to a finite system size. However, for the heat capacity peaks corresponding to the cluster to stripe and the stripe to bubble transitions, in Fig.~\ref{fig:heat_capacity_diff_box} we do see significant finite size effects. These peaks shift and become sharper and higher as the system size is increased. This might be seen as indicative that these are second order phase transitions, with a heat capacity divergence in the thermodynamic limit. However, recall that at a phase transition, in a small finite size simulation box the system fluctuates between the two phases. This leads to a double peak in the density histogram at that state point (or indeed the histogram of any other quantity that is a suitable order parameter for the transition). However, as can be seen in Fig.~\ref{fig:MC_prob_rho}, where we display the density histogram calculated at the value of $\mu$ corresponding to the peak in the heat capacity, there is a single peak {(the corresponding chemical potential values are $\beta\mu \approx 4.0$ and $\beta\mu \approx 3.1$ for $L=40\sigma$ and $L=60\sigma$, respectively)}. We obtain very similar distributions for state points either side of where the heat capacity peak occurs. {An alternative order parameter that is more sensitive to periodic ordering is the density Fourier mode amplitude,
\begin{equation}
	\left| \rho_{\vec{k}} \right|  = \sqrt{ \Big(  \sum_{j=1}^{N} \cos( \vec{k}\cdot\vec{r}_j) \Big)^2 + \Big(  \sum_{j=1}^{N} \sin( \vec{k}\cdot\vec{r}_j) \Big)^2 }.
\end{equation}
In Fig.~\ref{fig:MC_Prob_Rho_k} we display the histogram of $|\rho_\vec{k}|$ for the wavevector $\vec{k} = (k_{p},0)$, where $k_p\sigma=0.2\pi$, which is the value where there is a peak in $S(k)$. This order parameter histogram also has a single peak for values of $\mu$ where the heat capacity exhibits a peak.}

From the fact that there is only a single peak in {Figs.~\ref{fig:MC_prob_rho} and \ref{fig:MC_Prob_Rho_k}}, we infer that the transition from the cluster to striped state is simply a change in morphology, much like the micellisation process. We infer the same for the transition from the stripe to the bubble morphology. For low values of $(\beta\e)^{-1}$, we believe that the large heat capacity peak at the transition to the stripe phase and the strong finite-size effects are due to the fact that the stripes that are formed span the simulation box (see Fig.~\ref{fig:mc_snapshots}). The finite size box stabilises the stripes, damping some of the long wavelength fluctuations.

\section{Lattice DFT}
\label{sec:Lattice_DFT}

We now present results for  the structure and thermodynamics of the fluid, which are calculated using density functional theory, and compare with the MC simulation results. The mean-field DFT that we use is a generalisation of the theory presented in Ref.~\onlinecite{Hughes:2013aa} (see also references therein for other applications of the theory). The thermodynamic grand potential is approximated by
\begin{align}
 {\Omega} &=  k_B T \sum_{i=1}^M \left[\rho_i\ln(\rho_i) + (1-\rho_i)\ln(1-\rho_i) \right] \nonumber
         \\ 
         &+\frac{1}{2} \sum_{i,j}V_{i,j}\rho_{i}\rho_{j} + \sum_{i=1}^M (\Phi_i - \mu)\rho_i\,.
  \label{eq:avg_density_rho}
\end{align}
The equilibrium density profile is that which minimises ${\Omega}$, i.e.\ is the solution of
{
\begin{equation}
\frac{\partial{\Omega}}{\partial{\rho_i}} = 0,\,\, \text{for all}\,\, i.
\label{eq:min_cond}
\end{equation}

Thus, from Eqs.~\eqref{eq:avg_density_rho} and \eqref{eq:min_cond}} we obtain
\begin{equation}
  \rho_i = \( 1-\rho_i \) \exp \Big[ \beta \big(  -\sum_j V_{i,j}\rho_j - \Phi_i + \mu\big) \Big].
\end{equation}
This set of coupled equations are solved by Picard iteration.\cite{Hughes:2013aa} In order to make sure $\rho_i$ does not fall outside the interval $(0,1)$ during the iteration process, we introduce a mixing parameter, $\alpha$. The idea is that after each iteration, we mix the new density value with the previous one,
\begin{equation}
 \rho_i = \alpha \rho_i^\mathrm{new} + (1-\alpha)\rho_i^\mathrm{old}.
\end{equation}
The mixing parameter $\alpha$ typically takes a value in the range $(0.01,0.2)$. Too large {a} value of $\alpha$ leads to instabilities in the iteration, whilst if $\alpha$ is too small, it leads to slow convergence.

\subsection*{{DFT results and comparison with MC} }

\begin{figure*}[htbp]
  \centering
  \includegraphics[width=0.98\textwidth]{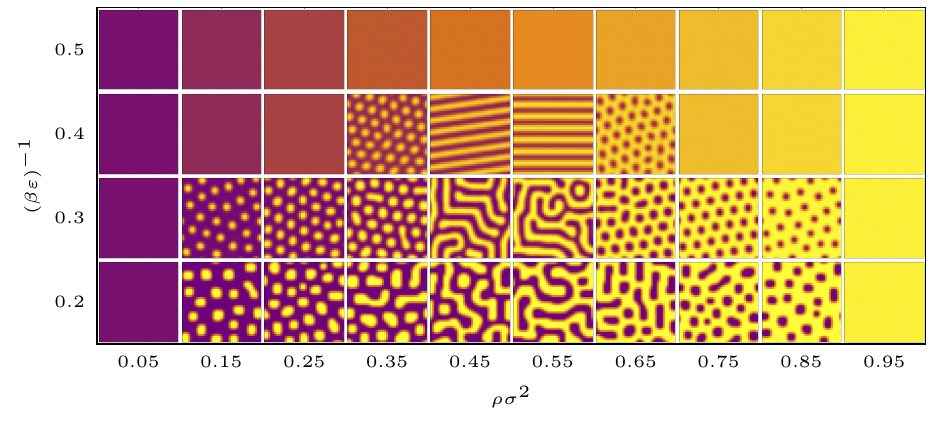} 
  \caption[Series of density profiles calculated from the lattice DFT]
  {A series of density profiles for varying values of $(\beta\e)^{-1}$ calculated using the lattice DFT for a $40\sigma\times40\sigma$ size system with random initial conditions, for $\beta A = 1.5$, $z_1=2$ and $z_2=0.2$ [c.f. Fig.~\ref{fig:mc_snapshots}]. The colours associated with each density value can be deduced from the top row of profiles, which are for $(\beta\e)^{-1} = 0.5$. }
  \label{fig:dft_snapshots}
\end{figure*}

In Fig.~\ref{fig:dft_snapshots}, we display examples of density profiles calculated using the lattice DFT for various values of the attraction strength parameter $(\beta\e)^{-1}$. These are obtained by initiating the Picard iteration with a flat density profile, to which is added a small amplitude random value at each lattice site. The density profiles show the same sequence of structures as observed in Fig.~\ref{fig:mc_snapshots} from the MC simulation, namely uniform, cluster, stripe, bubble and uniform as the chemical potential (density) is increased. The agreement between Fig.~\ref{fig:dft_snapshots} and Fig.~\ref{fig:mc_snapshots} is rather good. Within the DFT each of these different structures correspond to different solution branches of the grand potential. {The global minimum structure for a given state point contains no defects. Thus, in Fig.~\ref{fig:dft_snapshots} the vast majority of the structures displayed are not global minima of $\Omega$. To calculate the phase diagram, we calculate the free energy for defect-free structures, which are obtained by initiating the Picard iteration from profiles with the required structure, rather than from random initial conditions.} As $\mu$ is increased, there are points where these branches cross. At these points the solutions on the different branches have the same $\mu$, $T$ and pressure $p=-\Omega/V$, where $V=M\sigma^2$ is the area of the 2D system. Thus, the (incorrect) prediction from the mean-field DFT is that there are first order phase transitions between all the different structures.

We calculate the lines of thermodynamic coexistence in the phase diagram predicted by the DFT by selecting an initial lattice with a certain microphase separation and then change the chemical potential $\mu$ and follow that particular branch of solutions. For example, to find the coexistence curve for the gas to cluster transition, we start the DFT iteration with a uniform gas profile and increase $\mu$ with the new guess being the minimised density profile from the previous value of  $\mu$. While doing this we record the grand potential $\Omega$. Also, we start with an initial density profile corresponding to the cluster structure at a higher value of $\mu$ and then decrease $\mu$ following this branch of solutions. Coexistence is found when the pressure, temperature and chemical potential of the two structures are equal. The lines of coexistence define the boundaries in the phase diagram of where the different microphase separated structures occur.

\begin{figure}[htbp]
  \centering
  \includegraphics[width=0.45\textwidth]{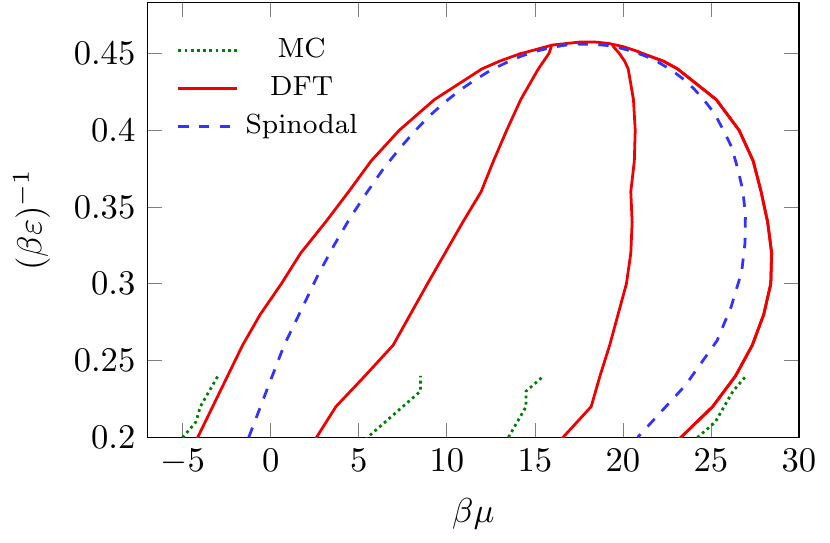} 
  \caption[Phase diagram with respect to chemical potential]{Phase diagram showing the instability threshold (spinodal, displayed as the blue dashed line) and the coexistence lines (red solid lines) obtained from DFT for varying values of the chemical potential $\mu$ and attraction strength $(\beta\e)^{-1}$. The location of the peaks in the heat capacity determined from the MC simulations for a $40\sigma\times40\sigma$ system are also shown, as the green dotted line. Note that these lines terminate where the peaks disappear (c.f.\ Fig.~\ref{fig:mc_snapshots}).}
  \label{fig:phase diagram}
\end{figure}

As shown in Fig.~\ref{fig:phase diagram}, we see that at the highest values of $(\beta\e)^{-1}$ (weak attraction) there is no microphase separation and the system exhibits a single uniform fluid phase. The DFT predicts microphase separation for values of $(\beta\e)^{-1}<0.45$. For the higher values in this range, e.g.~$(\beta\e)^{-1}=0.4$, the heat capacity from MC simulations in Fig.~\ref{fig:heat_capacity} has no discernible peaks. Nonetheless, comparing Fig.~\ref{fig:dft_snapshots} and Fig.~\ref{fig:mc_snapshots}, we see that the DFT is correctly predicting the structures formed, it is solely failing to describe the nature of the transition to the modulated structures. 

We also see a general shift of the occurrence of microphase ordering to higher values of $\mu$ as we increase $(\beta\e)^{-1}$. In Fig.~\ref{fig:phase diagram} we also display as green dotted lines the locations of the peaks in the heat capacity, from the MC simulations for a system of size $40\sigma\times40\sigma$. We see that these peaks lie close to the DFT coexistence lines for the gas to cluster transition and also the bubble to liquid transition. However, for the transitions to the stripe state, they are further away. We should emphasise, however, that these are subject to significant finite size effects. For a larger system, these are much closer to the DFT coexistence line.

The linear instability threshold line in Fig.~\ref{fig:phase diagram} is calculated numerically by starting from an initial density profile with the given average value of the density, but with small amplitude random fluctuations. We then determine whether the fluctuations grow over time as we iterate. The boundary of the region where they do grow is referred to as the spinodal in Fig.~\ref{fig:phase diagram}. We can also see that the instability line is completely inside the coexistence line. {An alternative (but entirely equivalent) way to calculate the spinodal is to determine when the uniform density solution to Eq.~\eqref{eq:avg_density_rho} ceases to be a mimimum. Consider a small amplitude harmonic density perturbation of the form
\begin{equation}
\rho_i=\rho+a e^{\mathrm{i}\vec{k}\cdot \vec{r}_i},
\label{eq:rho_harm}
\end{equation}
where the amplitude $a$ is a small parameter, $\vec{r}_i$ is the location of lattice site $i$ and $\vec{k}$ is any wavevector that is commensurate with the lattice. Substituting Eq.~\eqref{eq:rho_harm} into Eq~\eqref{eq:avg_density_rho} and then requiring that there is no solution except when $a=0$, is equivalent to the requirement that
\begin{equation}
\frac{1}{1-\rho}+\rho\beta V_d(\vec{k}) >0,
\label{eq:spin_cond}
\end{equation}
where $V_d(\vec{k})=\sum_jV_{i,j}e^{-\mathrm{i}\vec{k}\cdot \vec{r}_{i,j}}$ is the discrete Fourier sum of the potential, where $\vec{r}_{i,j}=\vec{r}_{i}-\vec{r}_{j}$. The quantity on the left hand side of Eq.~\eqref{eq:spin_cond} is equal to $1/S_{DFT}(\vec{k})$, where $S_{DFT}(\vec{k})$ is the static structure factor predicted by the DFT. Within the spinodal displayed in Fig.~\ref{fig:phase diagram}, Eq.~\eqref{eq:spin_cond} is no longer true for all $\vec{k}$ and thus the uniform density profile is no longer a minimum of the free energy.\cite{Archer:2008aa}
}

\begin{figure}[t]
        \centering
        \includegraphics[width=0.45\textwidth]{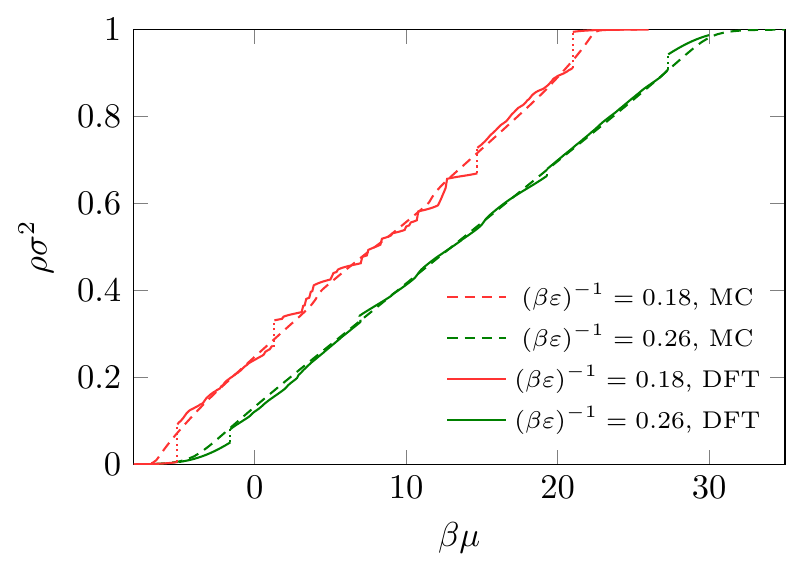}
        
        \includegraphics[scale=1]{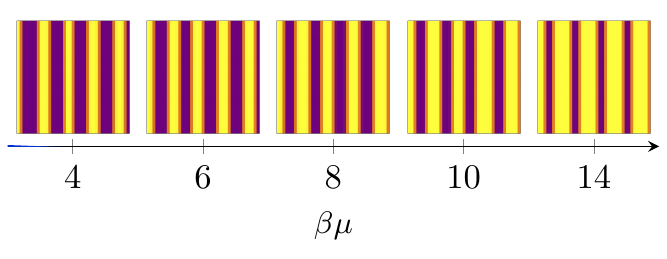}
        \caption{Top: a comparison of the average density as a function of $\mu$ for two different values of $(\beta\e)^{-1}$ from the MC simulations (dashed lines) and DFT (solid lines). The dotted line in the DFT curves show the jumps at which the transitions between the different morphologies occurs. Bottom: DFT density profiles showing the discontinuous changes in the stripes as we vary $\mu$ for fixed $(\beta\e)^{-1} = 0.18$, resulting in the non-smooth curves in the density plot above.}
         \label{fig:mc_dft_density}
\end{figure}

In Fig.~\ref{fig:mc_dft_density} we compare how the average density varies with chemical potential in the MC simulations with the results from DFT. We see that the MC simulation results show a smooth increase in the density. However, for sufficiently low values of $(\beta\e)^{-1}$, the DFT {gives jumps} in the density as we increase $\mu$. The jumps are plotted as dots in Fig.~\ref{fig:mc_dft_density}, which corresponds to the values of $\mu$ where microphase separation occurs. The magnitude of the jumps {decreases} as we increase $(\beta\e)^{-1}$. The jumps in the DFT {occur} because of various local minima in the free energy. Hence, the DFT has a tendency to stick to the initial density profile (local minimum) that we start from. Thus, the initial density profile is important for determining if the grand potential minimum that the iteration goes to is actually the global minimum. Different initial density profiles give us different local minima, which also depends on the box size, as expected. The DFT results are closer to the MC simulation results at higher values of $(\beta\e)^{-1}$ where there are more fluctuations in the system and the structural changes that occur in the system are smoother.

For example, when $(\beta\e)^{-1} = 0.18$ (typical of low values of $(\beta\e)^{-1}$), the DFT exhibits many discontinuities as we increase the chemical potential. This can be easily noticed in the middle portion of the curve in Fig.~\ref{fig:mc_dft_density} which corresponds to the {stripe} region. This is due to discontinuous changes in the width of the stripes that arise as we change the chemical potential. This is illustrated in the lower plots in Fig.~\ref{fig:mc_dft_density}, where we see that the width of individual stripes varies with changing chemical potential - i.e.\ not all stripes in Fig.~\ref{fig:mc_dft_density} have the same width. This confirms that the pattern formed is not necessarily the global equilibrium, since we expect the width of all the individual stripes to be identical at a global minimum.

\begin{figure}[t]
  \centering
  \includegraphics[width=0.45\textwidth]{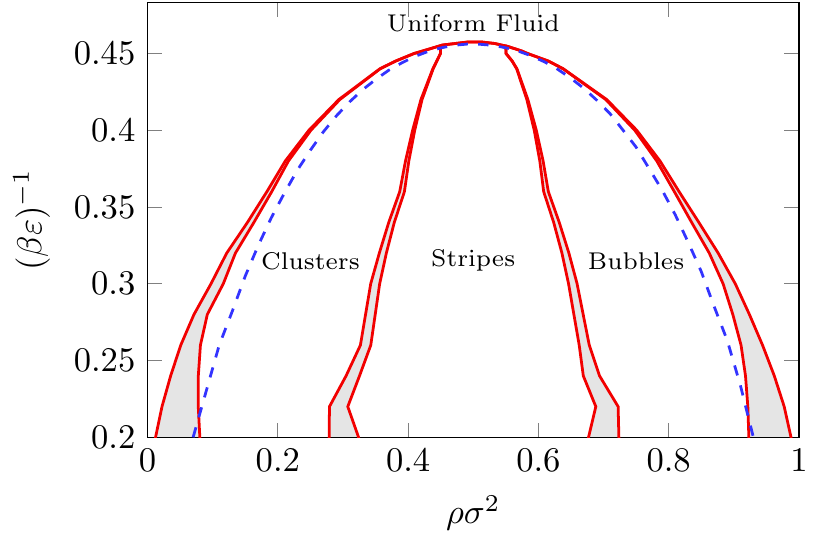} 
  \caption[Phase diagram with respect to density]
  	{Phase diagram showing the instability line (blue) and the coexistence lines (red) from DFT for varying values of the density $\rho$ and attraction strength $\e$, for fixed $\beta A = 1.5$, $z_1=2$ and $z_2=0.2$.}
  \label{fig:phase_diagram_2}
\end{figure}

Plotting the value of $(\beta\e)^{-1}$ at which the transitions occur as a function of density, we see that in this representation the phase diagram is symmetric around $\rho\sigma^2 = 0.5$ (see Fig.~\ref{fig:phase_diagram_2}). The instability line is fully within the region of the phase diagram where the uniform liquid is metastable. The shaded regions are the regions of coexistence between the two phases. We also see that the density range over which there is coexistence decreases as we increase $(\beta\e)^{-1}$.

\section{Continuum DFT approximation}
\label{sec:LS}

We now approximate the discrete lattice model by treating it with a continuum DFT, that enables a more straightforward calculation of quantities such as the linear instability threshold (spinodal) and other related quantities. This mapping from the lattice to a continuum assumes that the density profile $\rho_i$ varies slowly enough that we can treat it as a discretised representation of a continuous profile $\rho(\vec{r})$. This also enables us to convert the sums over lattice sites into integrals. Hence, the Helmholtz free energy $F=\Omega + \mu \langle N \rangle$ [cf. Eq.~\eqref{eq:avg_density_rho}], can be written as the following functional:
\begin{align}
	F &= \int f(\rho(\vec{r}))\,\mathrm{d}\vec{r} 
		+ \frac{1}{2}\iint \rho(\vec{r})\rho(\vec{r'})V(|\vec{r}-\vec{r'}|) \,\mathrm{d}\vec{r} \,\mathrm{d}\vec{r'}
	 \nonumber \\
	&	+ \int \rho(\vec{r}) \Phi(\vec{r}) \,\mathrm{d}\vec{r}
	\label{eq:LS:Helm_energy}
\end{align}
where $V( {r})$ is the pair potential in Eq.~\eqref{eq:tail}, $\Phi(\vec{r})$ is the external potential and $f$ is a local free energy per unit area given by
\begin{equation}
	f(\rho) = k_BT \left[ \rho \ln\( \rho \) + \( 1-\rho \) \ln\( 1-\rho \) \right] - \frac{\chi}{2} \rho^2\, .
\end{equation}
The first term is the free energy for a non-interacting ($\e=A=0$) lattice gas. The second term involving the parameter $\chi$ is a term to correct for the effect of the mapping from the lattice to the continuum, so that the continuum model gives the same free energy for the uniform fluid as the lattice model. The parameter $\chi$ is the following integrated difference between the continuum pair potential and the lattice potential:
\begin{equation}
	\chi = 2\pi  \int_\sigma^{r_c} rV(r) \, \mathrm{d}r - \sum_{<i,j>} V_{i,j}.
\end{equation}

The reason for mapping to a continuum model is that the following linear stability analysis is made somewhat more simple. The aim of the linear stability analysis is to determine where in the phase diagram the uniform fluid state becomes unstable, i.e. we locate the region of the phase diagram in which the microphase ordering occurs.

Consider a uniform fluid with density $\rho_0$. We wish to know whether any small amplitude density modulation will grow over time (fluid is unstable) or whether the amplitude will decrease (fluid is stable). Specifically we consider a density fluctuation of the form {[c.f.~Eq.~\ref{eq:rho_harm}]}
\begin{align}
	\rho	&= \rho_0 + \delta\rho(\vec{r},t)\nonumber
		\\[5pt]
		&= \rho_0 + \xi e^{\mathrm{i}\vec{k}\cdot\vec{r} + \omega t },
		\label{eq:LS:rho}
\end{align}
where $\xi$ is the initial amplitude of the sinusoidal perturbation that has wavenumber $\vec{k}$. The growth/decay rate of this mode is given by the dispersion relation $\omega = \omega(k)$, where $k=|\vec{k}|$.\cite{Archer:2004aa}

\begin{figure}[t]
  \centering
  \includegraphics[width=0.45\textwidth]{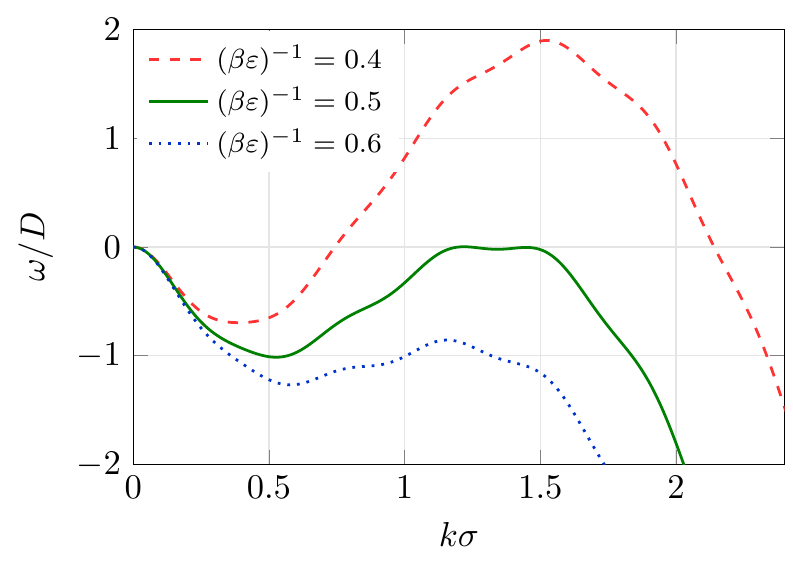} 
  \caption{The dispersion relation \eqref{eq:LS:omega} for varying attraction strength $\e$, for the uniform fluid with density $\rho\sigma^2=0.5$, for fixed $\beta A = 1.5$, $z_1=2$ and $z_2=0.2$.}
  \label{fig:disp_rel}
\end{figure}

To determine the time evolution of this non-equilibrium density profile, we require a theory for the dynamics of the colloids. This is supplied by dynamical density functional theory (DDFT), which shows that for Brownian colloidal particles the time evolution of $\rho(\vec{r},t)$ is governed by \cite{marconi1999dynamic, Archer:2004aa, Hansen:2013aa}
\begin{equation}
	\pdif{\rho}{t} = D \nabla \cdot \left[ \rho \nabla \cfrac{\delta \beta F}{\delta \rho}  \right],
	\label{eq:LS:dynamicaleq}
\end{equation}
where $D$ is the diffusion coefficient of the colloids. Note that for an equilibrium fluid, the chemical potential\cite{Evans:1979aa, Evans:1992aa, Hansen:2013aa}
\begin{equation}
	\mu = \frac{\delta F}{\delta \rho}
\end{equation}
is a constant. Thus, in Eq.~\eqref{eq:LS:dynamicaleq}, it is gradients in the chemical potential of the non-equilibrium fluid that drives the dynamics. Substituting Eq.~\eqref{eq:LS:rho} into Eq.~\eqref{eq:LS:dynamicaleq} together with Eq.~\eqref{eq:LS:Helm_energy} with the external potential $\Phi = 0$, and then linearising in $\delta\rho$, we obtain the following expression for the dispersion relation\cite{Archer:2004aa} {[c.f.~Eq.~\eqref{eq:spin_cond}]}
\begin{equation}
  \omega = -D k^2\( { \frac{1}{1-\rho_0} } - \beta \chi \rho_0 + \beta \rho_0 \hat{V}(k) \),
  \label{eq:LS:omega}
\end{equation}
where $\hat{V}(k)$ is the 2D Fourier Transform of the pair potential
\begin{equation}
  \hat{V}(k)= 2\pi \int_0^{\infty} rV(r) J_0(kr ) \, \mathrm{d} r,
\end{equation}
where $J_0(x)$ is the Bessel function of order 0. In Fig.\ \ref{fig:disp_rel} we display the dispersion relation for the uniform fluid with density $\rho\sigma^2=0.5$, for various values of $\e$.

\begin{figure}[htbp]
  \centering
  \vspace*{1em}
  \includegraphics[width=0.45\textwidth]{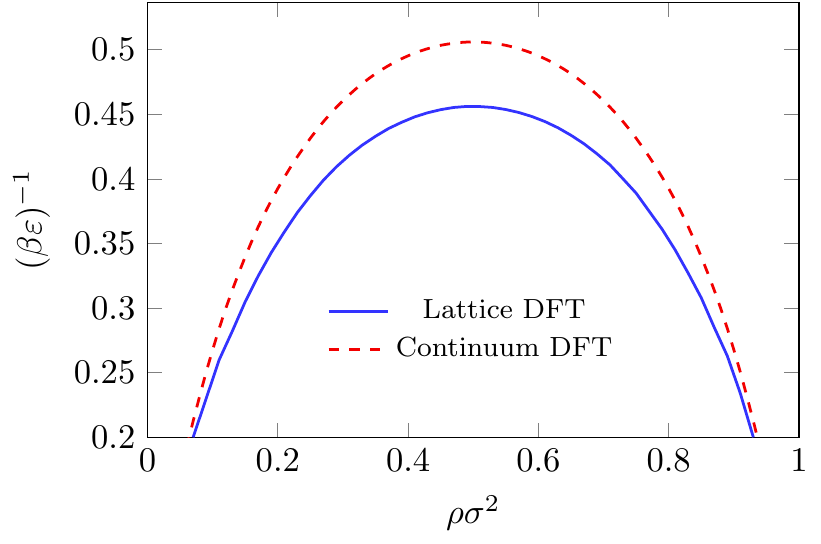} 
  \caption[Instability lines from lattice DFT and continuum DFT]{Phase diagram showing the linear instability threshold line for the lattice DFT (blue solid line) and also the instability line for the continuum DFT (red dashed line), calculated from dispersion relation in Eq.~\eqref{eq:LS:omega}.}
  \label{fig:instability_lines}
\end{figure}

From the dispersion relation, we can find the linear instability threshold line. Since we know that the system becomes unstable when $\omega > 0$, the instability line is calculated for values of $\e$ and $\rho_0$, where $\omega(k_c) = 0$, where $k_c$ is the value at which $\omega(k)$ is a maximum, i.e.
\begin{equation}
	\left. \dif{\omega(k)}{k} \right|_{k=k_c} = 0.
\end{equation}
The linear instability line is thus easily obtained from the dispersion relation and is displayed in Fig.~\ref{fig:instability_lines}. In this figure, we also display the linear instability line for the original lattice DFT model.

Comparing the two instability lines in Fig.~\ref{fig:instability_lines} shows that the maximum value of $(\beta\e)^{-1}$ where the system is linearly unstable is predicted to be a little higher in the continuum theory, {compared} to the lattice model. Comparing with Fig.~\ref{fig:mc_snapshots} and Fig.~\ref{fig:dft_snapshots}, we see that this simple calculation does indeed identify the region of the phase diagram where microphase separation is observed. Of course, it does not specify which structures (cluster, stripe or bubble) are formed, but it does allow one to narrow down to the relevant region of the phase diagram.

{We find the above analysis rather instructive: mapping from a lattice to continuum theory or vice-versa is a ``trick'' that is often performed to aid the analysis of a system. This procedure is clearly an approximation, but the fact that the two curves in Fig.~\ref{fig:instability_lines} are reasonably close to one another gives confidence that in the present situation the mapping is justified.}

\section{Conclusion}
\label{sec:conc}

In this paper we have studied a lattice model for 2D colloidal fluids where the colloids have attractive interactions at short separations, but repel at longer range. We model this by using a double-Yukawa pair potential between the particles. This SALR system self assembles to form different microphase separated structures. Using MC computer simulations and by calculating the heat capacity of the system as the chemical potential $\mu$ and the attraction strength coefficient $\e$ are varied, we determine where in the phase diagram the different morphology changes occur. At lower values of $(\beta\e)^{-1}$, the heat capacity exhibits peaks at the transitions between the different structures. The height of the peaks decrease as we increased $(\beta\e)^{-1}$, eventually disappearing. The peak at the transition from the gas to the cluster state and also for the bubble to liquid shows no system size dependence for systems greater than or equal to $40\sigma\times40\sigma$ in size. However, the peaks for the transitions to the stripe phase do change with system size, for the system sizes we were able to consider. By calculating how the number of lone particles {and the static structure factor} varies through the transition between the gas and the cluster phase, we conclude that this transition is a structural transition entirely akin to micellisation. The transition from the cluster to the stripe phase is very similar, except here occurring on a larger scale, by the gathering together of clusters to form stripes. This behaviour is also observed in living polymerisation, where a peak in the heat capacity is also observed.\cite{greer1998physical, DFD99, WD05}

Due to the fact that the pair potential \eqref{eq:DY} between the particles is fairly long ranged, the MC simulations can be computationally expensive. {Recall that we cut-off our slowly decaying potential at a range of $r_c=16\sigma$, which is much longer ranged than the potentials considered e.g.\ in Refs.~\onlinecite{Pekalski:2013aa, Pekalski:2014ab, Almarza:2014aa, selke1980two, landau1985phase}.} We only implemented the simple Metropolis MC algorithm, so correctly sampling for system sizes greater than $60\sigma\times60\sigma$ and for many state points was not feasible. To simulate efficiently for larger systems, a more sophisticated MC incorporating e.g.\ cluster moves is required. {This simple MC also limited what temperatures (i.e.~values of $(\beta\e)^{-1}$) we could go down to. For $(\beta\e)^{-1}=0.18$ we are confident that our MC simulations are correctly sampling the system. However, for lower temperatures, the algorithm struggles to sample a representative set of states in the time available. The low temperature properties of the model are interesting as it may be the case that at very low temperatures the structural transitions we observe become genuine phase transitions. It is certainly the case that other lattice models with competing interactions\cite{Pekalski:2014ab, Almarza:2014aa, selke1980two, landau1985phase} do exhibit phase transitions at low temperatures. We leave investigating this aspect to future work.}

We also used a simple lattice DFT to calculate density profiles for the system. Comparing Figs.\ \ref{fig:mc_snapshots} and \ref{fig:dft_snapshots}, the agreement between simulation and the mean-field DFT is rather good. The pair potential \eqref{eq:DY}, with the parameter values that we use, is fairly long ranged and slowly varying -- see Fig.~\ref{fig:potential}. In the case of purely attractive systems, when the pair potentials are long ranged and slowly varying (the classic mean-field situation) then one would not be surprised to find that mean-field DFT is accurate. However, given that the present system exhibits microphase ordering and is strongly fluctuating, it was not a-priori clear that the agreement between the DFT and the MC is as good as it is.

We also used the DFT to calculate the phase diagram and found that the heat capacity peaks in the MC simulations are close to the transition lines predicted by the DFT for the gas to cluster transition and the bubble to liquid transition. For the cluster to stripe and stripe to bubble transitions, they are somewhat further away. One aspect of the DFT is that at lower values of $(\beta\e)^{-1}$, the model exhibits many local free energy minima. This means that to use the DFT to calculate the phase diagram one needs to ensure one has a good choice of initial density profile. Starting from a density profile that is not good approximation, the iteration can go to a local minimum with a free energy value above that of the global minimum. Such behaviour is often observed in pattern forming systems. Thus, great care is required to determine the system sizes in which the system arranges in a state that is close in free energy value to the global minimum value.

Mapping the lattice model onto a continuum DFT yields a theory from which determining the linear instability threshold line using the dispersion relation is straightforward, enabling us to easily and rapidly determine the range of parameter values where the microphase ordering occurs. This provides a useful starting point if future analysis of the behaviour of systems with different pair potential parameter values is required.

\section*{Acknowledgements}
We would like to thank Mark Amos, Adam Hughes, Alberto Parola, Davide Pini and Nigel Wilding for helpful discussions at various stages of this work. BC and CC are both funded by EPSRC studentships.


%

\end{document}